\documentclass[fleqn,usenatbib]{mnras}

\usepackage{newtxtext,newtxmath}

\usepackage[T1]{fontenc}

\DeclareRobustCommand{\VAN}[3]{#2}
\let\VANthebibliography\thebibliography
\def\thebibliography{\DeclareRobustCommand{\VAN}[3]{##3}\VANthebibliography}

\usepackage{graphicx}	
\usepackage{amsmath}	
\usepackage{cite}
\usepackage{hyperref}
\usepackage{color}
\usepackage{bm}

\newcommand{\sgra}{Sgr A$^{*}$}

\defcitealias{Li2017}{L17}
\defcitealias{Yuan2009}{Y09}
\defcitealias{GRAVITY2018}{G18}

\title[\sgra\ flares]{A ``coronal-mass-ejection'' model for flares in Sagittarius A$^*$}

\author[Lin, Li \& Yuan]{
Xi Lin$^{1,2}$,
Ya-Ping Li$^{1}$\thanks{E-mail: liyp@shao.ac.cn}
and Feng Yuan$^{1,2}$\thanks{E-mail:fyuan@shao.ac.cn}
\\
$^{1}$Shanghai Astronomical Observatory, Chinese Academy of Sciences, Shanghai 200030, People’s Republic of China\\
$^{2}$University of Chinese Academy of Sciences, 19A Yuquan Road, Beijing 100049, People’s Republic of China}

\date{Accepted XXX. Received YYY; in original form ZZZ}

\pubyear{2023}

\begin{document}
\label{firstpage}
\pagerange{\pageref{firstpage}--\pageref{lastpage}}
\maketitle

\begin{abstract}
High-resolution near infrared observations with GRAVITY instrument have revealed rapid orbital motions of a hot spot around Sgr A*, the supermassive black hole in our Galactic center, during its three bright flares. The projected distances of the spot to the black hole are measured and seems to increase with time. The values of distance, combined with the measured orbiting time, imply that the spot is rotating with a super-Keplerian velocity. These results are hard to understand if the spot stay within the accretion flow thus provide strong constraints on theoretical models for flares. Previously we have proposed a ``CME'' model for the flares by analogy with the coronal mass ejection model in solar physics. In that model, magnetic reconnection occurred at the surface of the accretion flow results in the formation of flux ropes, which are then ejected out. Energetic electrons accelerated in the current sheet flow into the flux rope region and their radiation is responsible for the flares. In this paper, we apply the model to the interpretation of the GRAVITY results by calculating the dynamics of the ejected flux rope, the evolution of the magnetic field and the energy distribution of accelerated electrons, and the radiation of the system. We find that the  model can well explain the observed light curve of the flares, the time-dependent distance and the super-Keplerian motion of the hot spot. It also explains why the light curve of some flares have double peaks. 

\end{abstract}




\section{Introduction}

Sagittarius A$^*$ (\sgra), the super-massive black hole located at our Galactic center (SMBH; with mass $M \sim 4.1\times10^{6} \ {M_\mathrm{\odot}}$ and distance $D\sim 8.1 \ \mathrm{kpc}$), provides an excellent  laboratory to investigate extreme physics and accretion process due to its proximity (e.g., \citealt{Boehle2016, Gillessen2017, GRAVITY2018, GRAVITY2019, Ghez2008}; see reviews by \citealt{Melia&Falcke2001,2010RvMP...82.3121G,YuanNarayan2014}). It is one of the most under-luminous SMBH
with a bolometric luminosity in its quiescent state as low as $L_\mathrm{bol} \sim 5 \times 10^{35} \ \mathrm{erg}\ \mathrm{s}^{-1} \sim 10^{-9} \ L_\mathrm{Edd} $, where ${L_\mathrm{Edd}}$ is the Eddington luminosity. The accretion rate is as low as $(5.2-9.5)\times 10^{-9} \ {M}_{\odot} \mathrm{yr}^{-1}$ \citep{2000ApJ...545..842Q,Marrone2006,EHT2022a}, so it has been  an ideal target for studying hot accretion flows around black holes (\citealt{1995Natur.374..623N,1998ApJ...492..554N,Yuan2003,Yuan2004,2020ApJ...896L...6R,EHT2022a,EHT2022e}; see \citealt{YuanNarayan2014} for a review).

While Sgr A* spends most of its time in the quiescent state, it also exhibits multi-wavelength flares. These flares are most significant in the near infrared (NIR) and X-ray wavebands, which generally occur several times a day with a typical timescale of half an hour to an hour \citep{Dodds-Eden2011,Witzel2012,Neilsen2013,Li2015,Ponti2015,Yuan2016}. During the NIR flares, the flux density can be enhanced by a factor of 10 within $\sim 10$ minutes \citep{Genzel2003, Ghez2004, Trippe2007, Witzel2021}. X-ray flares are always associated with NIR flares \citep{2004A&A...427....1E,2006A&A...450..535E}, the luminosity of which is $L_{\mathrm{{3-79 \ kev}}} \sim 2 \times10^{35} \ \mathrm{erg s^{-1}}$, two orders higher than that of the quiescent state \citep{Baganoff2001,2003A&A...407L..17P,2008A&A...488..549P,Dodds-Eden2011,Eckart2013,Neilsen2013,Neilsen2015,Zhang2017, Boyce2019, Do2019,Haggard2019, GRAVITY2021}. Flares have also been observed at lower frequencies such as submm and radio wavebands, with some time delay with respect to the high frequency flares \citep{2003ApJ...586L..29Z,2006A&A...450..535E,Zadeh2006}. The high linear polarization degree ($\sim 20\%-40\%$) of the flares at NIR wavelengths suggests a synchrotron origin, but the emission mechanism for the X-ray flares is not fully conclusive. Many  models have been proposed,  the possible radiative processes invoked in the models include inverse Compton (IC) scattering of submillimeter photons, synchrotron-self-Compton (SSC) scattering of IR/
NIR photons, or pure synchrotron radiation from non-thermal electron \citep{2001A&A...379L..13M,Yuan2004,Zadeh2006,Broderick2006,Eckart2008,Dodds-Eden2009,Yuan2009,Dodds-Eden2010, Zadeh2012,Ponti2017, Subroweit2020}.  It is argued in some works that, although IC and SSC models can self-consistently explain the simultaneous IR/X-ray multiwavelength observations and the time delay between flares at high and low frequencies, synchrotron radiation model has the advantage that it can be achieved with physically plausible
magnetic field strength and electron density compared with IC and SCC models \citep{Dodds-Eden2009,Dodds-Eden2010}. 

A breakthrough for understanding Sgr A* flares was made by \citet{GRAVITY2018} (hereafter \citetalias{GRAVITY2018}), in which they reported the first detection of the continuous positional changes of Sgr A* during three prominent bright NIR flares with the near-infrared GRAVITY-Very Large Telescope Interferometer (VLTI) beam-combining instrument. The position centroids exhibit a nearly circular trajectory around the central black hole with a velocity of $\sim 0.3c$ and an orbital period of tens of minutes. During these flares, polarization angle also exhibits continuous rotation, presenting a roughly same period as that of the centroid motion, which is also consistent with the timescale of the flares. The centroid motion and polarization angle rotation can be well fitted by a radiating ``hot spot'' rotating around the black hole with its distance to the black hole being several $r_g$ in the projected plane. 

It is not trivial to quantitatively explain the GRAVITY observational results, as emphasized by, e.g.,  \citet{Matsumoto2020}.   The centroids trace 3/4 of a full loop in 30 min and this time corresponds to a Keplerian orbit radius of $\sim 7 r_g$ \citepalias{GRAVITY2018}. But this radius is smaller than the detected distance between centroids and their medians most of the time, which is $\sim$(10-20)$r_g$ (refer to Figure 1 in \citet{GRAVITY2018} and Figure 1 in \citet{Matsumoto2020}). In another word, the hot spot has to rotate with a super-Keplerian velocity to match both orbit radius and period. Therefore  a "hot spot" confined on the equatorial plane of an accretion flow  is hard to explain the results, because the theory of hot accretion flow which predicts that the flow should have a sub-Keplerian azimuthal velocity \citep{YuanNarayan2014}. Furthermore, the centroid trajectory of the July 22 flare shows a gradually increasing orbital radius (refer to Figure 1 in \citet{GRAVITY2018} and Figure 1 in \citet{Matsumoto2020}), which also indicates the "hot spot" should not stay in the accretion flow otherwise the radius would decrease as it is advected to the black hole with the accretion flow. 

Recently many models have been proposed to investigate the nature of flares based on the (GR)MHD simulation of black hole accretion flows, and these models often invoke magnetic reconnection \citep{Guo2014,2016ApJ...818L...9G,Sironi2014, Werner2016, Werner2018} occurred in the accretion flow as the mechanism of electron acceleration \citep[e.g.,][]{2020MNRAS.495.1549N,Nathanail2022,Petersen2020,Zhao2020,2020MNRAS.497.4999D,Chatterjee2021,Porth2021,Scepi2022,2022ApJ...926..136W,2022ApJ...937L..34K,Ripperda2022}. The radiation of these energetic electrons accelerated in the magnetic reconnection events is then responsible for the observed ``hot spot''.  
However, with a few exceptions as we will discuss in Section \ref{comparison}, none of these works has presented detailed calculations of radiation and comparisons with observations especially with the GRAVITY observations.  

An additional important clue to the nature of flares comes from the radio observations carried out by VLA at 43 and 22 GHz \citep{Zadeh2006}. A radio flare was detected at these two frequencies, with the peak flare emission at 43 GHz leading the 22 GHz flare by $\sim 20-40$ minutes. Such phenomenon is a strong evidence for the ejection and expansion of a radio-emitting plasma blob from the accretion flow during the flare activity. In fact, such kind of physical association between flares and episodic ejection has been observed widely in other black hole sources, such as radio galaxy 3C120 \citep{Marscher2002, Olmstead2008,Chatterjee2009,Casadio2015},  blazar PKS 1510–089 \citep{Park2019}, black hole X-ray binary GRS 1915+105 \citep{Mirabel1994, Fender1999, Jones2005}, low-luminosity AGN M81 \citep{King2016} and M87 \citep{Hada2014}. Although it is possible that not all flares are associated with ejection of blobs, these observations suggest that a physical model of flares should at least be able to provide a framework of simultaneously explaining the associated ejection. 

\begin{figure*}
	\includegraphics[width=1.\textwidth]{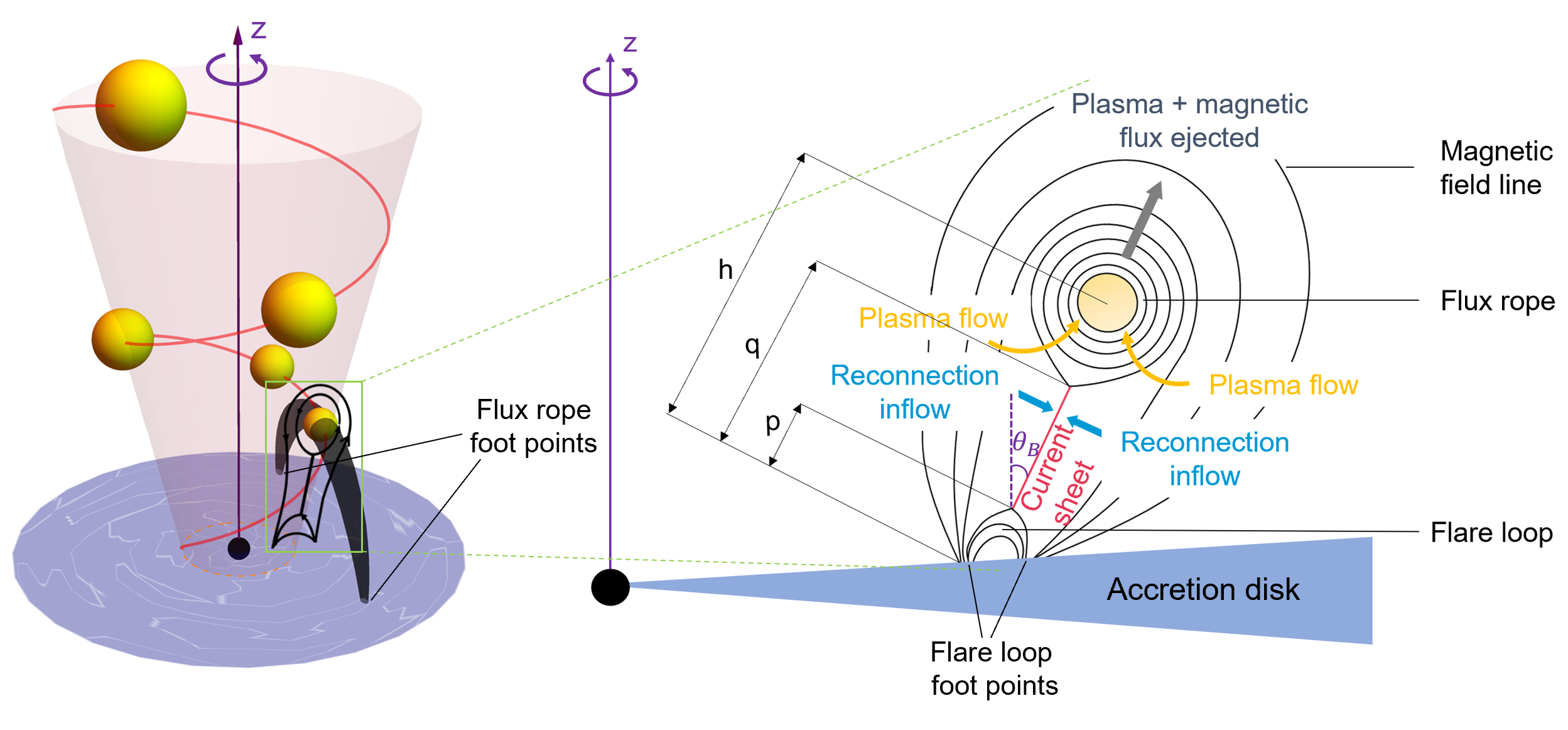}
    \caption{A schematic illustration of our model (left panel) and a zoomed-in plot of the flux rope in its 2D cross section (right panel). In the left panel, the main radiation region is from the center of the flux rope where it is represented by a yellow hot spot. The helical trajectory of the hot spot is depicted by the red curve and the hot spots in different times are plotted. Note that the hot spot is expanding as it is ejected outward. In the right panel, the flux rope is ejected outward in a fixed angel $\theta_B$ inclined with respect to the normal direction of the equatorial plane (i.e., $z$-direction). The loop and the flux rope are connected by the current sheet (marked by the red line).}
    \label{fig:schematic}
\end{figure*}

Stimulated by this important observational result, \citet{Yuan2009} (hereafter \citetalias{Yuan2009}) have proposed a model to interpret the flares and the associated ejections of blobs. The model is by analogy with the standard model of coronal mass ejection (CME) on the Sun \citep{Lin2000}, since CME is often associated with solar flares, which is very similar to the black hole sources as we have mentioned above.  The schematic figure of the system is shown in Figure 2 of that paper (see also Figure \ref{fig:schematic} in the present work). There exist closed magnetic field lines in the coronal region of the accretion flow, which could be emerged out of the main body of the accretion flow due to Parker instability. The turbulent and differential rotation of the accretion flow twist the field lines and result in reconnection and the formation of flux ropes in the coronal region, similar to the formation of prominence in the solar corona. The reconnection enhances the magnetic pressure and reduces the magnetic tension force. Consequently, the flux ropes are ejected out by the  magnetic pressure force. 

This scenario has been confirmed recently by the detailed analysis to the three dimensional GRMHD simulation data of black hole accretion flow  \citep{Miki2022}. The accelerated electrons in the current sheet will flow into the flux rope and the flare loop regions through the reconnection outflow, as shown by Figure \ref{fig:schematic}. The radiation of these energetic electrons  explains the observed flares, while the ejected flux rope is responsible for the observed ejected blobs.  \citet{Li2017} (hereafter \citetalias{Li2017}) have applied the \citetalias{Yuan2009} model to Sgr A* and successfully interpreted the NIR and X-ray light curves of Sgr A* flares by calculating the dynamics of the ejected blobs and the evolution of the energy distribution of energetic electrons accelerated by the magnetic reconnection process. 

In the present work, we investigate whether we can explain the new observational results from \citetalias{GRAVITY2018} based on the framework of \citetalias{Yuan2009} and \citetalias{Li2017}. 
The structure of the paper is as follow. In Section \ref{sec:2} we calculate the dynamic properties of hot spots based on \citetalias{Yuan2009} and \citetalias{Li2017}. We then evolve the non-thermal electrons for the radiation calculation injected from the reconnection region in Section \ref{sec:3}. The ray-traced calculation is described in Section \ref{sec:4}. We show the results and compare with observations in Section \ref{sec:5}. We discuss and summarize our results in Section \ref{sec:discussion} and Section \ref{sec:6}, respectively.

\section{Dynamics of the flux rope}
\label{sec:2}

\subsection{The basic scenario of the Model}

The original \citetalias{Yuan2009} model is two-dimensional. To calculate the radiation, we need to convert this two-dimensional model to three-dimension, which is shown schematically by the left plot of  Figure \ref{fig:schematic}. 
We start our calculations from a flux rope newly formed due to magnetic reconnection. The flare loop stays on the equatorial plane, keeping a constant distance $R_{\mathrm{lp}}$ away from the black hole and following a sub-Keplerian rotation velocity of the underlying RIAFs \citep{YuanNarayan2014}:
\begin{equation}
    \Omega_{\mathrm{lp}}= k\Omega_{\mathrm{0}},
\end{equation}
where $0  \textless k  \textless 1$ is the proportionality factor, and $\Omega_{\mathrm{0}}$ is the Keplerian angular velocity at $R_{\mathrm{lp}}$ \citep{Bardeen1972}:
\begin{equation}
    \Omega_{\mathrm{0}} = \frac{M^{1 / 2}}{R_{\mathrm{lp}}^{3 / 2} + a M^{1 / 2}},
\end{equation}
where $M$ and $a$ are the mass and the dimensionless angular momentum of the black hole. 

The ejected flux rope follows a spiral motion since it has angular momentum, its trajectory is shown in the left panel of Figure~\ref{fig:schematic} by the yellow sphere and red line. The motion of the flux rope can be decomposed into two components, i.e. the azimuthal (or toroidal) and poloidal motions. The inclusion of the toroidal motion of the ejected flux rope is one of the main improvements we make in this work for the dynamics of the flux rope compared to \citetalias{Yuan2009} and \citetalias{Li2017}, where the toroidal motion is not considered for simplicity. The poloidal motion of the flux will be calculated  in Section \ref{dynamics}. Note that the poloidal motion of the ejected flux rope has an inclination angle $\theta_B$ with respect to the normal direction of the accretion flow, as shown in the figure. This is because the poloidal motion of the ejected flux rope has to follow the direction of the large-scale poloidal magnetic field, which is inclined with respect to the accretion flow in this way, as indicated by our GRMHD numerical simulations \citep{Miki2022}. As we will see later, this will explain  the  increasing distance of the hot spots observed by GRAVITY \citepalias{GRAVITY2018}. 

In our model, the rotation of the ejected flux rope is super-Keplerian.  The reason is as follows. From the GRMHD simulations by \citet{Miki2022}, we find that the flux rope  has the same angular velocity as its footpoint, i.e., the flare loop, which is  $\Omega_{\mathrm{rp}}=\Omega_{\mathrm{lp}}$, within a distance of $\sim 60 r_g$ from the black hole.  This distance marks the Alfvén surface where Alfvén speed $V_A (\equiv B/\sqrt{4\pi \rho})$ is equal to the wind speed. 
Physically, this is because the strength of the field is so strong that it can resist bending caused by differential rotation when the field line is close to the central black hole; the field line thus acts as a rigid rod that rotates around the black hole with the same angular velocity. Plasmoids enveloped in the flux rope co-rotate with the field line  due to the flux frozen-in. 
Due to the constant angular velocity of the ejected flux rope but increasing distance to the rotation axis, the plamoids will thus possess a larger angular velocity compared to the the local Keplerian velocity as it is ejected outward, i.e., its motion is super-Keplerian. We note that this is in good consistency with the GRAVITY observations \citepalias{GRAVITY2018} and emphasized by \citet{Matsumoto2020}.

Beyond this Alfvén surface, the field strength decreases to a certain extent that magnetic tension cannot effectively suppress differential rotation. Therefore the magnetic field lines start to be more twisted and eventually become toroidal-component dominated. 
In this region (e.g., $r\gtrsim60r_g$), the angular velocity of the flux rope is calculated instead according to the conservation of angular momentum in Newtonian frame, i.e., 
\begin{equation}
    \Omega_{\mathrm{rp}}=\Omega_{\mathrm{lp}}r_0^2/r(t)^2,
\end{equation}
where $r_0=60r_g$. 

\subsection{The poloidal motion  of the ejected flux rope}
\label{dynamics}

The detailed calculation of the poloidal motion of the flux rope can be referred to \citetalias{Yuan2009} and \citetalias{Li2017}. Here we only present a brief description. 
The dynamics of the ejected flux rope is controlled by three forces, namely the magnetic pressure, magnetic tension, and gravity. Due to the reconnection, the magnetic pressure force will become significantly larger than the sum of magnetic tension and the gravity. 
As a result, the flux rope will be ejected out by the magnetic pressure force. Magnetic dissipation in the current sheet will convert the magnetic energy to heat and accelerate electrons in the current sheet, and these electrons will be transported to the flux rope and the flare loop region located above and below the current sheet, respectively, through the reconnection outflow. 

\begin{figure}
	\includegraphics[width=\columnwidth]{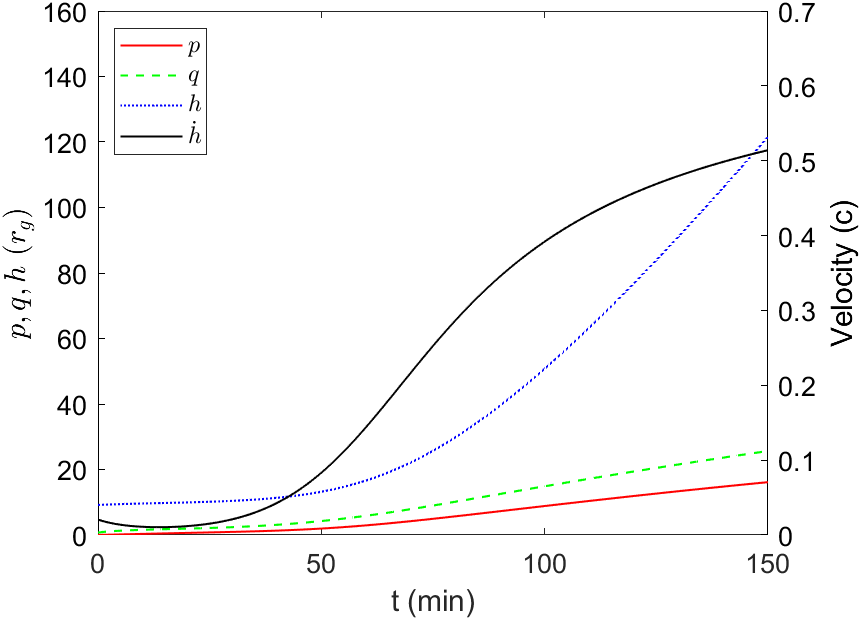}
    \caption{Dynamical evolution of the flux rope. The flux rope can be ejected out to a height of $h \sim 110 r_g$ with velocity of 0.5 speed of light within 150 minutes after the reconnection occurs.}
    \label{fig:1}
\end{figure}

The two foot points of the flux rope embedded in the magnetic arches are anchored at the surface of the accretion flow.  We approximately treat $\pi L_0$ as the total length of the magnetic arch, which should increase with the eruption of the flux rope. In this work, we take it as a constant and $L_0 = 50\ r_{\rm g}$ as in \citetalias{Li2017} for simplicity. The effective length of the magnetic arch should be determined by the length of the arch when the emission from the flux rope is strongest.  During the flares, the flux rope can be ejected outward ($h\gtrsim50\ r_{\rm g}$) as we show below (i.e., Figure~\ref{fig:1}), which would give a large effective $L_{0}$ of $\sim50\ r_{\rm g}$. Note that there is some degeneracy between the value of $L_0$ and other model parameters so we don’t treat it as a free parameter.   The central part of the flux rope has a tilted height $h$ (see the right panel of Figure \ref{fig:schematic}).


The explicit illustration of the system is shown in the right panel of Figure \ref{fig:schematic}. The dynamical evolution of the flux rope is depicted by five quantities, namely the ``height'' of bottom and top tips $p$ and $q$ of the current sheet, the ``height''\footnote{Here we use quotes mark because $\theta_B\neq 0$.} of the flux rope center $h$, the poloidal velocity of the flux rope $\dot{h}$ and the total mass enveloped in the flux rope $m$. These five quantities can be deduced by solving the equation of motion 
\begin{equation}
    m \gamma_{\mathrm{b}}^{3} \frac{d^{2} h}{d t^{2}}=\frac{1}{c}\left|\boldsymbol{I} \times \boldsymbol{B}_{\mathrm{ext}}\right|-F_{\mathrm{g}},
\end{equation}
where $\gamma_{\mathrm{b}}=1 / \sqrt{1-\dot{h}^{2} / c^{2}}$ is the Lorentz
factor of the flux rope, $\boldsymbol{I}$ is the integrated current intensity  inside
the flux rope, $\boldsymbol{B}_{\mathrm{ext}}$ is the total external magnetic field (apart from $\boldsymbol{B}_{\mathrm{I}}$ generated by $\boldsymbol{I}$ itself) measured at the center of the flux rope, and $F_{\rm g}$ is the gravity from the central SMBH. The first term on the right-hand side depicts magnetic force and the second term depicts the gravitational force. 

Following \citetalias{Yuan2009} and \citetalias{Li2017}, in order to obtain an analytical solution of the motion of the flux rope, and by analogy with the magnetic field in the solar surface, the spatial distribution of the magnetic field is assumed to be described by 
\begin{equation}
B(\zeta)=\frac{2 i A_{0} \lambda\left(h^{2}+\lambda^{2}\right) \sqrt{\left(\zeta^{2}+p^{2}\right)\left(\zeta^{2}+q^{2}\right)}}{\pi\left(\zeta^{2}-\lambda^{2}\right)\left(\zeta^{2}+h^{2}\right) \sqrt{\left(\lambda^{2}+p^{2}\right)\left(\lambda^{2}+q^{2}\right)}},
\label{eq:2}
\end{equation}
where $\zeta=r+i z, A_{0}=B_{0} \pi \lambda_{0}$ is the source field strength and $\lambda=5r_g$ is the
half-distance between the two field line foot points of the flare loop. Note that the magnetic field is time-dependent due to the evolution of $p,q$ and $h$. We emphasize that this distribution of magnetic field is a weakness of our model. We should in principle adopt the realistic magnetic field, which is, however, very uncertain. We expect that the main effect will be on the prediction of the polarization, as we will discuss in Section~\ref{polarization}.  

There are seven free parameters in this model: (i) The distance of the flare loop center from the central black hole $R_{\mathrm{lp}}$. We choose it to be $6.5 r_g$, where $r_{g}$ is the gravitational radius of the SMBH\footnote{\citet{Miki2022} find that only flux ropes formed beyond $10-15 r_g$ can be ejected out while those formed inside this radius usually stay within the accretion flow. In our fiducial model, the flux rope is formed at $h\sim 10 r_g$ although its ``footpoints'' are at $6.5 r_g$ (refer to Figure \ref{fig:1}). This corresponds to its  spherical radius of $\sim 12 r_g$, thus the flux rope can be ejected out. }. This is in the favored range given by \citet{GRAVITY2020a}. We note that our model is not sensitive to the exact value of $R_{\mathrm{lp}}$. (ii) The proportionality factor $k$, which is set as $k=0.93$. That means the rotation velocity is sub-Keplerian as predicted by the theory of RIAFs \citep{YuanNarayan2014}. (iii) The tilted angle between the direction of the poloidal motion  of the flux rope and spin axis. We set this angle $\theta_B = 173^{\circ}$ (or $\theta_B = 7^{\circ}$). 
This small tilted angle indicates the ejected blobs are highly collimated, similar with jets. (iv) The background magnetic field $B_0$ in the vicinity of the reconnection region. We choose $B_0=120 \ \mathrm{G}$, which is a factor of a few larger compared to the average magnitude of 10-50 Gs (e.g.,\citealt{Yuan2003,Dodds-Eden2010,Boyce2019,EHT2022e}). This is reasonable given that the magnetic field has strong spatial fluctuation and  we focus on strong flares in highly magnetized regions where the magnetic field should be greater than the average value. (v) The electron number density in the inner accretion flow region $n_{\mathrm{e}} = 1.75\times 10^7\ \mathrm{cm^{-3}}$ and (vi) the density ratio $\xi=10$ for the flux rope with respect to the background density $n_{\mathrm{e}}$. (vii) The Alfv\'{e}n Mach number $M_{\mathrm{A}}$ defined as the reconnection inflow speed divided by the local Alfv\'{e}n speed near the reconnection regions,  determining the efficiency of the reconnection. In this work we set $M_{\mathrm{A}}=0.5$, which is slightly higher than the predicated values from the magnetic reconnection theory \citep{Liu2017,Guo2020}. Note that the adopted parameters above should not be regarded as a unique solution for modeling the NIR flares as we did not fully explore the parameter dependence in the current work due to the complexity of the radiative transfer calculations as described below.  We note that the values of these main parameters adopted in the present paper are basically the same as those in  
\citetalias{Li2017}. In this way, we hope to investigate whether our original 
\citetalias{Li2017} model, with  extension of including more consistent general relativity radiative transfer calculations, can also explain the new \citetalias{GRAVITY2018} results. These parameters together with which will be introduced in non-thermal electron evolution and  ray-tracing calculation are summarized in Table \ref{tab:1}. 

\begin{figure}
	\includegraphics[width=\columnwidth]{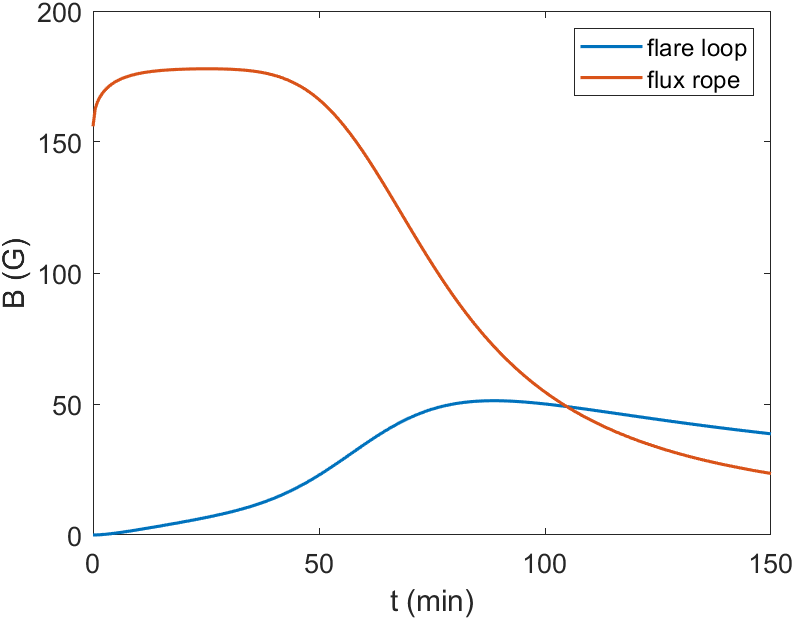}
    \caption{Evolution of the magnetic field in the flare loop and ejected flux rope. }
    \label{fig:2}
\end{figure}

As the system evolves, energy stored in the magnetic field releases and transforms to other forms of energy, e.g., the kinetic energy of the bulk motion of the flux rope, and energy of the heated and accelerated electrons. Figure \ref{fig:1} shows the evolution of $p,q,h$ and $\dot{h}$. During the whole evolution stage, the size of flare loop, which is determined by the lower tip of current sheet $p$, increases slowly ($\lesssim 10\ r_{\rm g}$). This indicates that the expansion of the flare loop region is insignificant and it is justified that the loop always stays on the equatorial plane. By contrast, the evolution of $h$, which measures the height of the flux rope, is dramatic, indicating the eruptive process of episodic ejection. In our fiducial model, the flux rope is formed at $h \sim 10r_{\rm g}$ or $r \sim 12r_{\rm g}$ in spherical coordinates. We found that if the initial electron density $n_\mathrm{e} \cdot \xi$ in the flux rope is too large, or the magnetic field magnitude in the reconnection region $B_0$ is too weak or the newly formed flux rope is too close to the black hole, the catastrophic eruption would fail and the flux rope would no longer be ejected out. This phenomenon is ubiquitous in solar observations. It is interesting to mention that similar scenario is also found in GRMHD simulation \citep{Miki2022}. They found that at small radii, inside of $\sim 10-15 r_{\rm g}$, few flux ropes can be ejected out, likely due to the strong gravitational force of the black hole which is similar to the absence of wind at small radii. As shown by Figure \ref{fig:1}, the flux rope is ejected out over 100 $r_g$ and reaches a velocity of $\sim 0.5 c$  within 150 min after the eruption occurs. This value is similar to  the ejection speed obtained in GRMHD simulations \citep{Miki2022}. The mean size of the flux rope can be roughly expressed by ($h-q$), which is $\sim 100 r_g$ after 150 min. But we will see that such a large flux rope in the later stage leave no observational signatures due to its strong cooling. The length of the current sheet ($\sim q-p$) is of the order of several $r_{\rm g}$, again consistent with that found in the GRMHD simulations \citep{Miki2022}.
The evolution of the spatially-averaged magnetic field within the flux rope and flare loop during the reconnection process is shown in Figure \ref{fig:2}. It can be seen that the magnetic field in the flux rope drops rapidly, which means the reconnection proceeds efficiently owing to the high $M_{\rm A}$. 

\begin{figure}
	\includegraphics[width=\columnwidth]{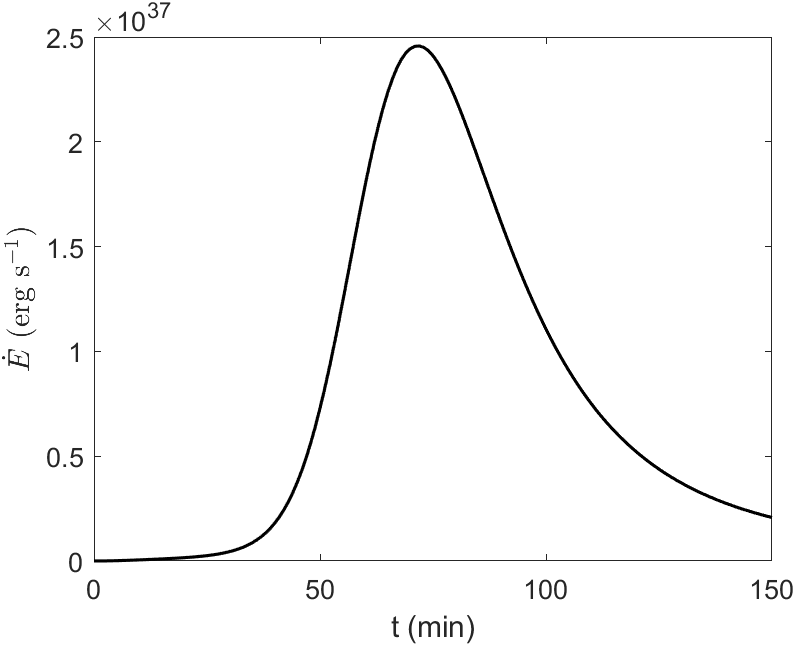}
    \caption{Total energy release rate during the magnetic reconnection in the model. The observed timescale of a typical flare event is about one hour.  }
    \label{fig:3}
\end{figure}

The electromagnetic energy release rate during the reconnection is given by
\begin{equation}
\dot{E}(t)=\pi L_{0} S(t),
\end{equation}
where $S(t)\propto E(t) \times B(t) $ is the Poynting flux in the current sheet. The reconnection event shows a peak power $\dot{E}_{\mathrm{{max}}} \sim 2.5 \times 10^{37}$ erg $\mathrm{s^{-1}}$ with a typical timescale of about one hour, as shown by Figure \ref{fig:3}, which is controlled by Alfv\'{e}n and reconnection time scales. This is comparable to the typical timescale of flares in \sgra.  As a consequence, the total energy released in this process is $  \textgreater 10^{40}$ erg, 
much larger than the estimation of $\ \gtrsim 10^{38}$ erg limited by two observed simultaneous near-infrared and X-ray flares \citep{Ponti2017}. This shows that there is enough energy available to reconnection to account for the observed energy released during the flares.
 
\section{Evolution of non-thermal electrons accelerated by reconnection}
\label{sec:3}
\begin{figure}
	\includegraphics[width=\columnwidth]{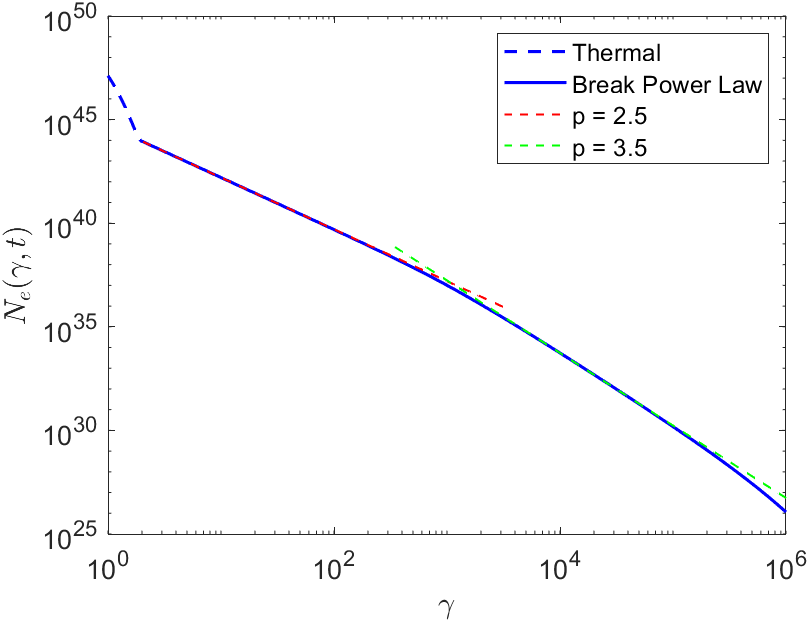}
    \caption{Electron energy distribution for the flare loop at $t = 120$ min after the reconnection starts. The electron temperature $T_{\mathrm{e}}=3.4\times 10^9 \ \mathrm{K}$, and mean magnetic field $B=45.2\ \mathrm{G}$. The distribution can be divided into three parts, i.e., the thermal part denoted by the dashed blue line, and two power-law parts with power-law indexes being $p=2.5$ and $p=3.5$, transited at  $\gamma \sim 10^3$. }
    \label{fig:4}
\end{figure}

Magnetic reconnection is well recognized as a high efficient way to energize electrons from thermal pool to high energy non-thermal electrons. The  acceleration process is found to be related to the ambient plasma-$\beta$ and magnetization $\sigma$ \citep[e.g.,][]{Ball2018}. In the present paper, we assume $10\%$ of the released energy  is delivered to non-thermal electrons, which is quite plausible in the low-$\beta$ magnetically-dominated region of hot accretion flow surface based on the previous theoretical and simulation works \citep{Ball2018,Chatterjee2021}. 
Particle-in-cell (PIC) simulations of magnetic reconnection indicate that the accelerated electrons follow a power-law distribution \citep{Guo2014,Sironi2014,Werner2017,Werner2018}, which can be written as
\begin{equation}
\dot{N}_\mathrm{pl}(\gamma)=\dot{n}_{\mathrm{pl}} \gamma^{-p_{\mathrm{e}}}, \quad \gamma_{\min } \leq \gamma \leq \gamma_{\max },
\end{equation}
where $\dot{n}_{\mathrm{pl}}$ is the non-thermal electron injection rate, $p_{\mathrm{e}}$ is the power law index, $\gamma_{\mathrm{min}}$ and $\gamma_{\mathrm{max}}$ are the low- and high-energy cutoffs of the distribution. The explicit value of $p$ is still not well determined, which could range from 1.0-2.5 depending on many factors like plasma-$\beta$ and magnetization $\sigma$ \citep[see, e.g., ][]{Guo2020,Li2021}. Observations show that the infrared spectral index of Sgr A* is $\alpha=-0.6 \pm 0.2\left(F_{\nu} \propto \nu^{\alpha}\right)$ \citep{Hornstein2007,Witzel2018}, which is expected for pure synchrotron radiation emitted from a power law energy distribution electrons with $p_{\mathrm{e}}=1-2 \alpha \sim 2.2$. The X-ray spectral index, on the contrary, has a large uncertainty. The value ranges from $-1.5 \pm 0.3$ to $0.5_{-1.3}^{+0.9}$ (\citealt{Porquet2003,Baganoff2001,Baganoff2003}). Then the inferred power-law index could be as steep as $p_{\rm e}\sim4.0$ if the X-ray comes from a population of single power-law electrons. However, the electron cooling timescale due to synchrotron losses is $ t_{\text {cool }}=1.29 \times 10^{12} \nu^{-1 / 2} B^{-3 / 2} \mathrm{~s}$, which is less than 10 mins for $B\sim 50G$ and $\gamma \sim 10^3$. The synchrotron cooling thus should be important in shaping the single power-law function into the broken power law.  This is also suggested by the steep spectral
index between IR and X-ray wavelengths of $\alpha_{\rm K-X}\sim-1.1$ to $<-1.5$.
Based on these, we set the power law index $p_\mathrm{e}=2.5$ for the initial injected electrons, which is also in agreement with the analytical studies of the first order Fermi process in current sheets which usually predicts  $p_{\mathrm{e}} \sim 2.5$ \citep{Pino2005}. We choose the maximum Lorentz factor of electrons $\gamma_{\mathrm{max}}=10^6$. Note that, as long as $\gamma_{\mathrm{max}}$ is high enough (e.g., $\gtrsim 10^4$), the value of $\gamma_{\mathrm{max}}$ is not so important when we focus on the NIR flares with $p_{\mathrm{e}}>2$ as considered here. The values of $\gamma_{\mathrm{min}}$ and $\dot{n}_{\mathrm{pl}}$ can be determined jointly by assuming the thermal and non-thermal electrons are smoothly connected at $\gamma_{\mathrm{min}}$ with an injected fraction $10\%$ of thermal energy from reconnection event, as shown in Figure~\ref{fig:4} (see \citetalias{Li2017} for details).

These accelerated non-thermal electrons flow into the flare loop and the flux rope. However, how the electrons distribute in each region are unclear and non-trivial. In this work, we simply assume that half of the electrons flow into each region. Given that cooling will be important if high-energy electrons contribute significantly to the NIR flares, we take into account  cooling due to the expansion of flaring region and due to synchrotron radiation \citep{Dodds-Eden2010, Li2017, Petersen2020,Chatterjee2021,Scepi2022}. We evolve the electron energy spectrum by considering injection as well as cooling according to the following continuity equation \citep{Blumenthal1970}:
\begin{equation}
\frac{\partial N_{\mathrm{e}}(\gamma, t)}{\partial t}=Q_{\mathrm{inj}}(\gamma, t)-\frac{\partial[\dot{\gamma} N_{\mathrm{e}}(\gamma,t)]}{\partial \gamma}.
\end{equation}
The escaping of non-thermal electrons from flaring regions are neglected. The detailed discussion of this equation can be referred to \citetalias{Li2017} and \citet{Dodds-Eden2009}. Figure \ref{fig:4} shows the energy spectrum of the flare loop at $t = 120$ min. There is clearly a break in the power law distribution around $\gamma \sim 10^3$; the power-law index beyond this $\gamma$ transits from $p=2.5$ to $p^\prime =3.5$. Because the radiation at NIR is likely from the electrons around this break, it is noted that a careful treatment of the electron distribution, i.e. a broken power law instead of a single power law distribution, should be taken.

\section{Calculation of radiative transfer}
\label{sec:4}

To calculate the radiation of the system, we employ the publicly available numerical code GRTRANS \citep{Dexter2009,Dexter2016}, which provides a self-consistent fully relativistic ray tracing radiative transfer calculation. As in most works, a ``fast light'' treatment is adopted to neglect the light travel time. Therefore, the direct emission and lensed emission will reach the camera at the same time. We note that this simplification may impose some modifications on the image and flux at each snapshot, and thus on the light curves, centroid motion and polarization as well \citep{Younsi2015, Ball2021}. 


In Section \ref{sec:3}, we have deduced the total non-thermal electron number $N_{\mathrm{e}} (\gamma,t)$ at each time in the flare loop and flux rope respectively. We further assume that these nonthermal electrons within the flux rope and flare loop regions satisfy a Gaussian distribution \citep{Broderick2006}
\begin{equation}
n_{\mathrm{S}(r,t)}=n_{\mathrm{S}}^{0} \exp \left[-\frac{\lvert \Delta r\rvert^2}{2 R_{\mathrm{0}}^{2}}\right],
\label{eq:density}
\end{equation}
where $\Delta r \equiv r-r_{\rm S}$ is the displacement from the region (hereafter ``spot'') center, and $R_{\mathrm{0}}$  controls the size of the spot. As the acceleration and spatial distribution of non-thermal electrons may depend on the microphysics of reconnection \citep[e.g.,][]{Guo2020}, i.e., the magnetization parameters, which is also hard to determine from the first principle for our semi-analytical modeling, we thus for simplicity fix $R_{\mathrm{0}}=2 \ r_g$ in both regions to be consistent with $R_{\mathrm{0}} \lesssim2.5 \ r_g$ favored by \citet{GRAVITY2020a}. The electrons are confined in $R_s = p \sim$ several $r_g$ for the flare loop and $h-q \sim$ tens of $r_g$ for the flux rope. We have to notice that the initial shape of the flux rope should not be spherical but arched (see left plot of Figure \ref{fig:schematic}). However, as the eruption proceeds, the complete structure of the flux rope cannot be well maintained. The flux rope will soon be entangled because its foot points are anchored in the different radii of the accretion flow where the angular velocities are different. This entanglement would exacerbate kink stability which could further disintegrate the end parts of the flux rope and finally only the main central part would be preserved. Therefore it is reasonable to adopt a spherical hot spot model for simplicity. The number density at the spot center $n_{\mathrm{S}}^{0}$ can be determined by integrating the distribution function, together with Eq. \ref{eq:density}
\begin{equation}
    \int^{R_S}_{0}4\pi r^2 n_{\mathrm{S}}dr = N_e.
\end{equation}

The configuration and magnitude of the magnetic field is described by Eq. \ref{eq:2}. It only has a poloidal component $\left(B_{r} \text { and } B_{\theta}\right)$. As suggested by \citetalias{GRAVITY2018}, a poloidal field is preferred to explain the rotation of polarization angle. In realistic situations, field configuration can be much more complicated. On the one hand, accretion and rotation can twist the field lines. On the other hand, the turbulence driven by magneto-rotational instability (MRI) can cause the field disordered. This effect is out the scope of our current work and we leave it for our future work based on 3-D GRMHD simulation of \citet{Miki2022} where magnetic field is self-consistently evolved. But as discussed before, a pure poloidal magnetic field is still an acceptable simplification 
given that the magnetic field is dominated by the poloidal component inside the Alfvén surface.

The radiative transfer is calculated in the Kerr metric and the physical quantities used to fulfil the calculation have been transformed into Boyer-Lindquist coordinates. More details about how to transform the velocity and magnetic field into four-vector forms can be referred from Appendix \ref{Appendix}. We choose the same coordinate systems to describe the motion of the hot spots as was adopted by \citet{Matsumoto2020} (refer to Figure 2 in their paper).
Dimensionless spin of the black hole $a=0.9$ and our results are not sensitive to its exact value. We choose this value because it agrees with the parameter survey conducted by \citet{GRAVITY2020a} and \citet{EHT2022e}. The fiducial inclination angle, defined as the angle between the line of sight and the normal direction of the accretion flow, is set to be $i=173^{\circ}$ to explain the apparent circular centroid motion and to ensure the clockwise motion of the spot. The almost face-on configuration is also favored by both GRAVITY \citep{GRAVITY2020a} and EHT \citep{EHT2022a,EHT2022e} results, which give $180^{\circ}\ga i \ga 140^{\circ}$. The longitude of the ascending node $\Omega$ is chosen to match both centroid trajectory and light curves in the flare epoch. Two types of ejected flux rope are considered, with one being ejected  towards the observer (forward model, $\theta_B=173^\circ$) while another being ejected away from the observer (backward model, $\theta_B=7^\circ$). In both cases, the brightness centroid exhibits apparent clockwise motion in the projected sky since their rotation direction follows the accretion flow. The field-of-view of the camera is [200,200] $\mu as$ in both x and y directions with a numerical resolution of  [1,1] $\mu as$ in each direction. For other parameters, the mass of the black hole $M = 4.1\times10^{6} \ {M_\mathrm{\odot}}$ and the distance from us is $8.1 \ \mathrm{kpc}$. 

With these setups, the polarized synchrotron radiation and images at NIR K-band at different time epochs from accelerated non-thermal electrons obtained from Section\ \ref{sec:3} are calculated using the GRTRANS code. The results are then compared with observations from \citetalias{GRAVITY2018}. Because of the large parameter space and non-triviality of radiative transfer calculations, we do not resort to any rigorous statistical method like Markov Chain Monte Carlo to fit the observational data. Instead, we manually adjust some additional geometrical model parameters combined with the model parameters adopted in the dynamical evolution stage until our model can match salient features of \citetalias{GRAVITY2018} observations.  The effect of some model parameters will be explored to diagnose their respective roles on different aspects of the results and some of them have been discussed in \citetalias{Li2017}.

\begin{table*}
 \caption{Main parameters of fiducial model. 'f' and 'b' represent forward and backward moving flux rope respectively. And the ray-tracing parameters for the flare loop are as same as the forward flux rope. }
 \label{tab:1}
 \begin{tabular}{cccccccc}
 \hline \hline Dynamics & $R_{\mathrm{lp}}(r_g)$ & $\theta_B$ (f/b) & $k$ & $B_0 (G)$ & $n_{\mathrm{e}} (\mathrm{cm^{-3}})$ & $\xi$ & $M_A$ \\
 
 \hline \ & 6.5 &$173^\circ$/ $7^\circ$ & 0.93 & 120G & $ 1.75\times 10^7  $ & $10 $ & $0.5$ \\
 \hline \hline Electron evolution & $p_{\mathrm{e}}$ & $\gamma_{\mathrm{max}}$ & $L_0 (r_g)$\ & $R_0 (r_g)$ & & & \\
 \hline \ & 2.5 & $10^6$ & 50 & 2 & & &\\ 
 
 \hline \hline Ray-tracing & $a$ & $i$ & $\phi_0$ (f/b) & $\Omega$ (f/b) &$M ( {M_\mathrm{\odot}})$ & $D$ (Kpc) & \\
 \hline \ & 0.9 & $173^{\circ}$ & $256.4^{\circ}$/$285^{\circ}$ & $270^{\circ}$/$110^{\circ}$ &  $4.1\times10^{6}$ & $8.1$ & \\
 \hline \hline
\end{tabular}
\end{table*}

\section{Results}
\label{sec:5}

\subsection{Light curves}
\begin{figure}
	\includegraphics[width=\columnwidth]{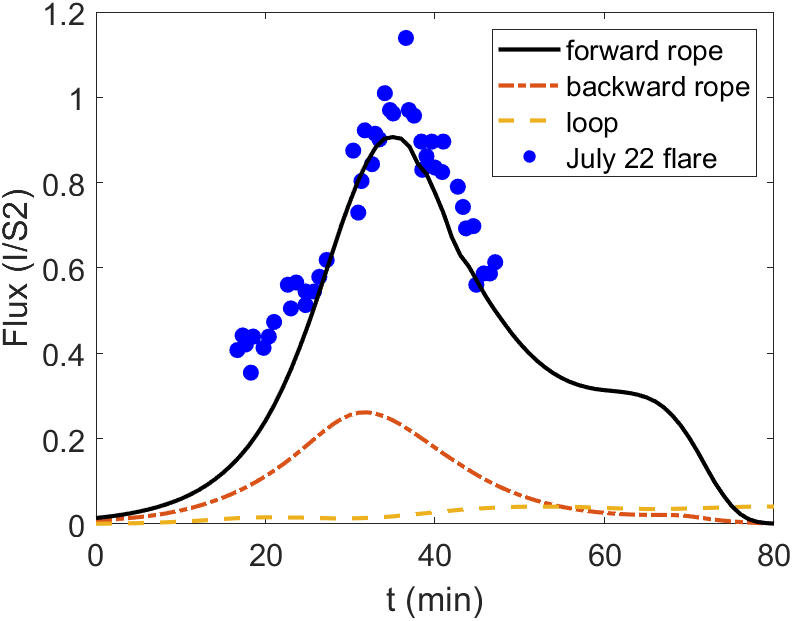}
    \caption{Light curves produced by the forward 
    and backward moving flux ropes and the flare loop. 
    The blue dots represent the observation data of the July 22nd flare from \citetalias{GRAVITY2018}. The flux is in units of the flux of S2 ($K_S = 14$, or 15 mJy). The starting time of our model is shifted to match the observational data. Similar treatment is made to the following figures.}
    \label{fig:lightcurve}
\end{figure}

Figure \ref{fig:lightcurve} shows the light curves produced by different regions and their comparison with observations.  The flare loop, and the forward and backward moving flux rope of our fiducial model are represented by different lines,  and the observational results of July 22 flare from \citetalias{GRAVITY2018} are shown as blue dots. The emission from the forward flux rope can well explain the main features of the observed light curve, including the timescale and luminosity. We can see from the figure that the emission of the flux rope region dominates over the flare loop by more than a factor of ten and shows much stronger variability than the flare loop.
There are several reasons for this enhancement. Firstly,  the energetic electrons in the flux rope get more concentrated in the high magnetic field region. Secondly, both the Doopler beaming effect  due to the high ejection/rotation speed (to the forward flux rope) and the general relativistic lensing effect (to the backward flux rope) can amplify the emission of the flux rope. As we will discuss later in Section~\ref{sec:discussion}, the former effect tends to be stronger. 
It should be noted that this is based on the assumption that we re-distribute energy and number of the non-thermal electrons to two flare regions equivalently. Results could be different if we inject a larger fraction of non-thermal electrons into the flare loop. By contrast, \citetalias{Li2017} finds the emission from the loop and flux ropes are comparable, which is because the effects mentioned above are not considered there. 

In our following analyses, we attribute the observed hot spot observed by \citetalias{GRAVITY2018} to the forward ejected flux rope rather than the flare loop and backward flux rope since the radiation of the forward flux rope can fit the light curve better. The flux rope moving toward us (forward flux rope) shows about 4-5 times stronger emission than the rope away from us (backward flux rope). This is mainly due to the beaming effect which can be in analogy with jets and counter jets. Although the emission from the  current backward flux rope does not fit the observational data, such a result may be parameter dependent. we suspect that under some model parameters, the general relativistic lensing effect may be able to more strongly amplify the radiation of the backward flux rope  and result in a stronger flare than the forward flux rope. 
In addition, we will see in the next section that the backward flux rope can also reproduce the hot spot trajectory. Hence we cannot fully rule out this possibility. 

\subsection{Centroid motion of the hot spot}

One of the most important results from GRAVITY observations is the centroid motion of the hot spot during the \sgra\ flares. The July 22 flare exhibits an apparent clockwise circular motion. The centroid position for the other two flares analysed in \citetalias{GRAVITY2018} does not form an obvious loop pattern due to the large uncertainties but still presents counter-clockwise rotating trend. Hence we focus on the orbit motion of the July 22 flare. The continuous positional changes exhibit $\sim 4/3$ of a full loop over 30 mins, corresponding to an orbit period of $\sim 40$ mins. The medians of the centroid are coincident with the location of the black hole determined from the mass center of the S2 orbit. 

\begin{figure}
	\includegraphics[width=\columnwidth]{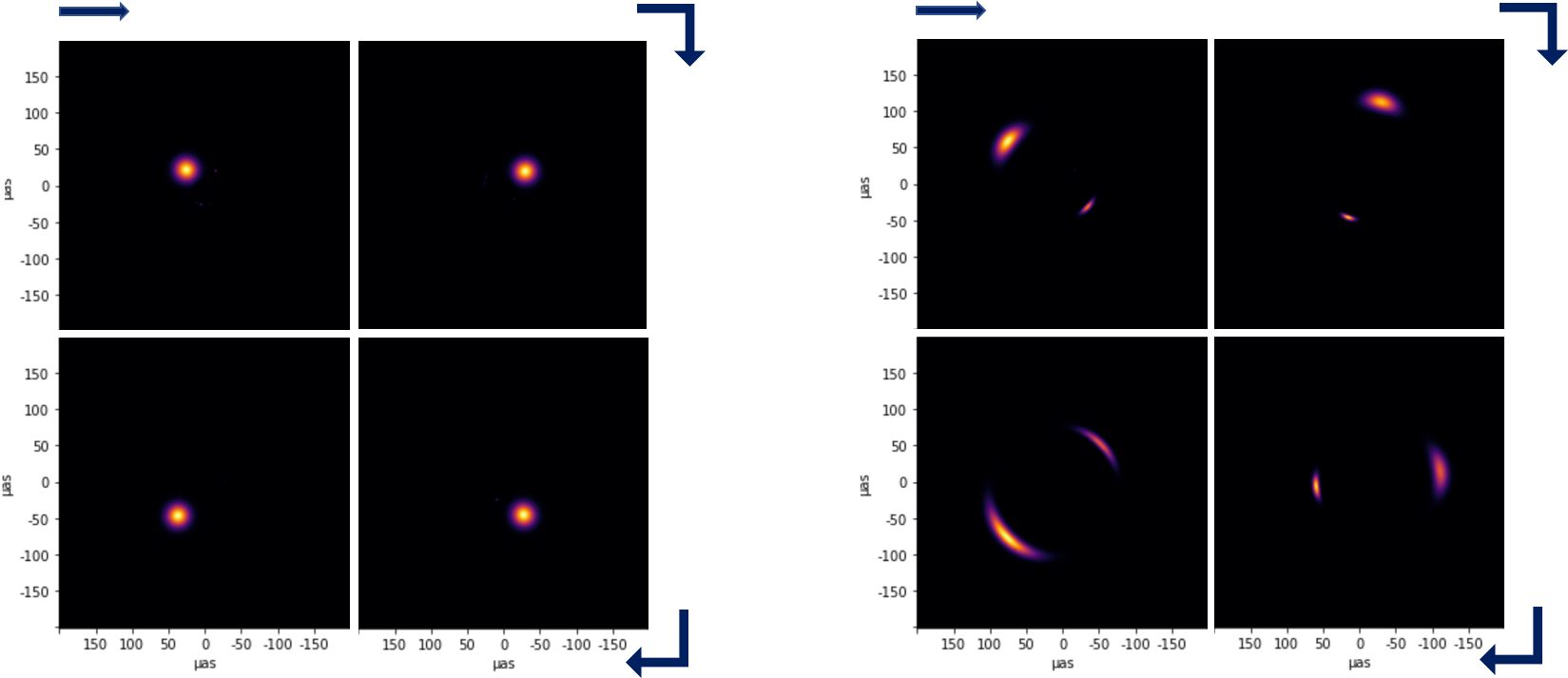}
    \caption{Ray-tracing images of the forward (left panel) and backward (right panel) moving blobs. The time interval between two adjacent panels is 10 mins. For the backward blob, two bright spots appear on the camera screen due to strong gravitational lensing, which has noticeable influence on the brightness centroid position. When the blob comes closer to the back side of the black hole, the lensing effect becomes stronger and the bright spot begins to be present as bright arcs as shown in the bottom left panel. Note that we do not consider any observational beam effect.}
    \label{fig:12}
\end{figure}

Figure~\ref{fig:12} shows the ray-tracing images of the forward and backward moving flux rope. One compact hot spot appears in the projected sky for the forward flux rope while two bright spots appear on each side of the black hole for the backward model. This is simply due to the strong gravitational lensing effect.

\begin{figure}
	\includegraphics[width=\columnwidth]{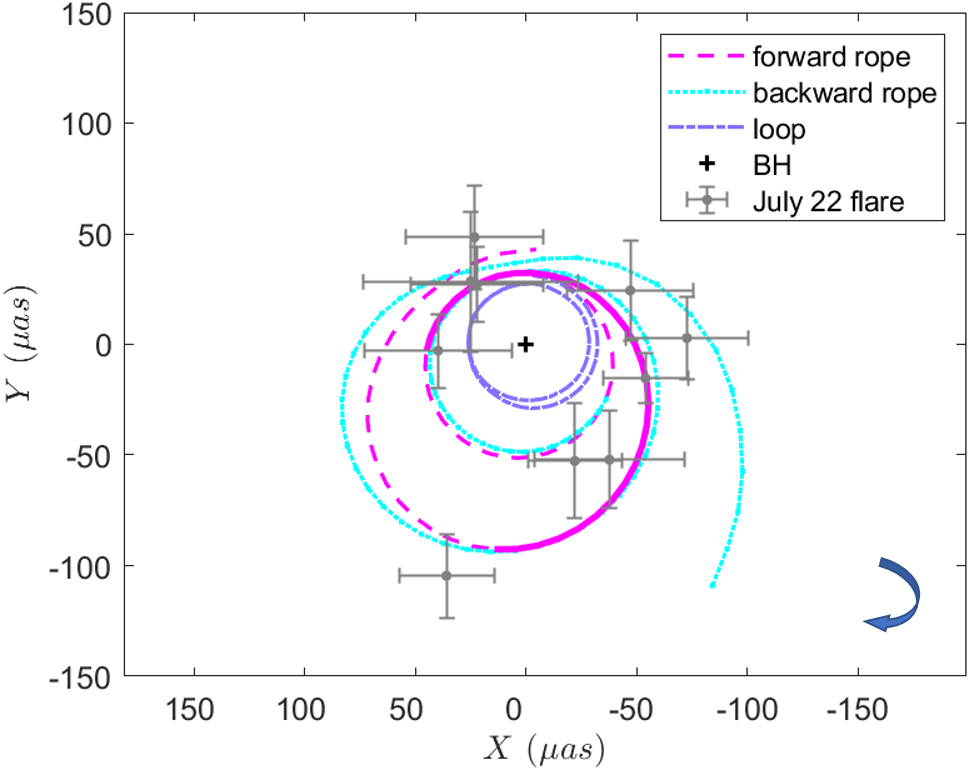}
    \caption{Centroid motion of the hot spot and its comparison with the July 22nd flare result. The 80 mins trajectory (in accordance with Figure\ \ref{fig:lightcurve}) of the flux rope is plotted and 30 min trajectory of the forward flux rope during the flare is thickened by the solid magenta line. The arrow in the bottom right indicates that these spots exhibit clockwise looped motion on the camera plane.}
    \label{fig:11}
\end{figure}

Figure \ref{fig:11} shows the centroid motion of the two flux ropes and the flare loop obtained by our ray-tracing calculations\footnote{We reasonably assume that usually only one flux rope exist during a flare, but the flare loop of course always exists. In this case, the observed hot spot should be the radiation flux-weighted combination of the flux rope and the flare loop, which is dominated by the flux rope.}, together with their comparisons with observations. The exact locations of each component are determined by the flux-weighted calculation. The helical motion of the ejected flux ropes in our model is represented as orbital motion in the projected plane.  On the projected plane, the trajectories of the two flux ropes are helices (magenta and cyan lines in Figure \ref{fig:11}), while that of the flare loop is nearly a circle (purple line in Figure \ref{fig:11}). The 30-min trajectory of the forward flux rope during the flare time is highlighted by the solid magenta line. 
Both the forward and backward flux ropes can satisfactorily explain the centroid motion of the NIR flares \citepalias{GRAVITY2018}. A remarkable feature is the increasing projected distance over time that fits with observations. More importantly, the ejected flux rope scenario with spiral motion makes the blob maintain its high angular velocity attained at small radius as it voyages to larger radius, which can then explain the observed super-Keplerian motion of the hot spot. 

From the Figure \ref{fig:11} we can rule out the flare loop model because its orbital radius is too small to be compatible with the observation. The orbit lies almost entirely inside the observation data points and the distance (Figure \ref{fig:11}) to the black hole is $\sim$ 3 times smaller than the last observed centroid. While we can enlarge the orbital radius by changing model parameters, it is difficult for the flare loop to show any helical motion as suggested by the observations. 
We find that the orbit of the flare loop in our model does not complete a closed orbit either. The reason is that our magnetic field and non-thermal electron distribution are evolving with time
and the emission region is extended in space rather than a point source. Therefore, the emission-weighted centroid may shift from the original location after one orbit.

\subsection{Polarization}
\label{polarization}

\begin{figure*}
	\includegraphics[width=0.8\textwidth]{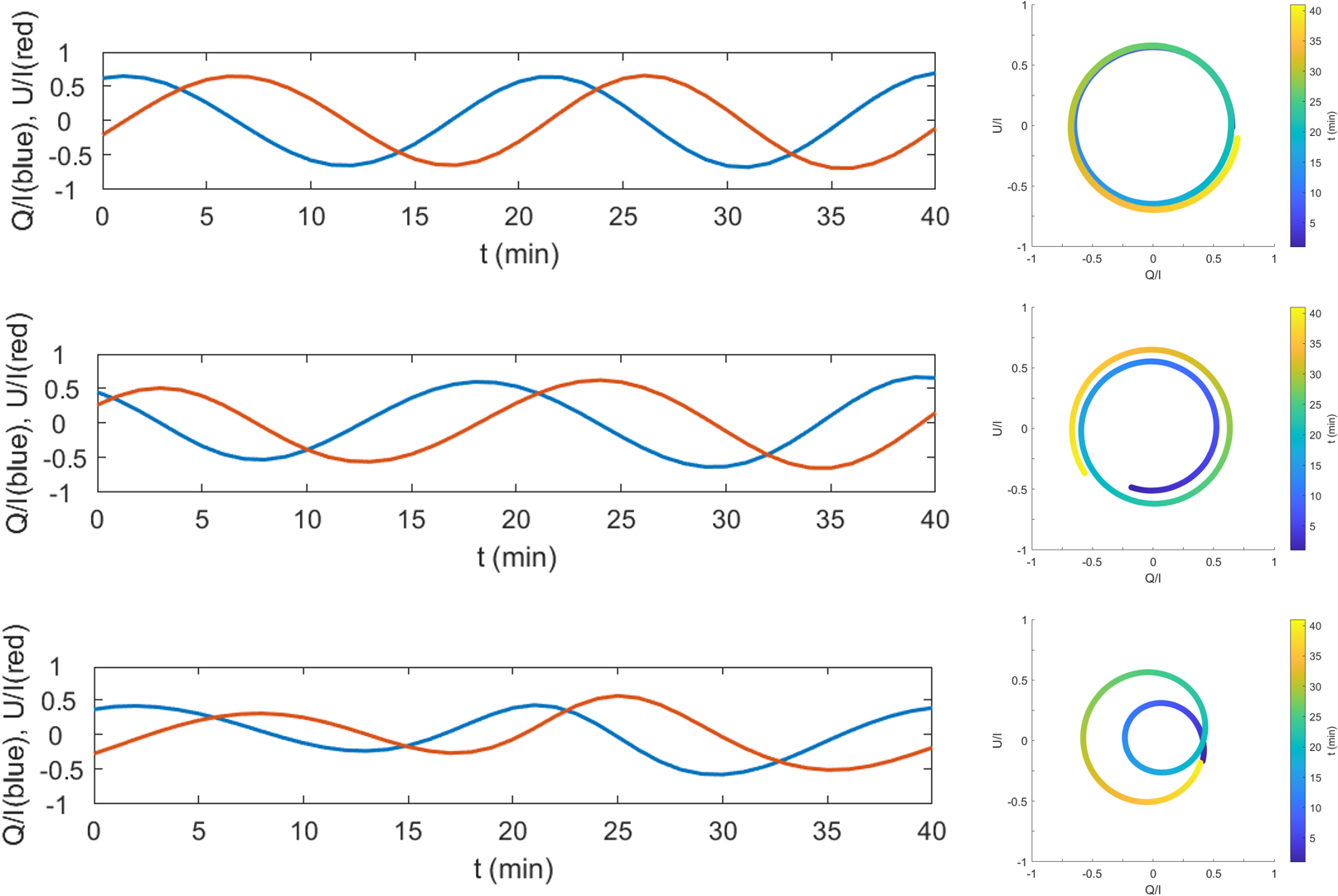}
    \caption{Polarized emission from the forward flux rope (up panels),  backward flux rope (middle panels) and flare loop (lower panels). By doing so, the Stokes parameters $I, Q, U$ are integrated over the entire source region. The polarization period of all model is $\sim$ 20 min, approximately a half of the orbit period. }
    \label{fig:15}
\end{figure*}

The polarization results are very sensitive to the configuration of the magnetic field. As we have emphasized before, the spatial distribution adopted in our model is not that realistic. 
But it is still worthwhile to investigate the polarization signatures of our model and compare them with observations.  

The evolution of polarization angle and fraction from our model is shown in Figure \ref{fig:15}. We plot the polarization $Q/I$ and $U/I$ in one orbital period, i.e., 40 mins. 
The July 28 flare shows a rotation of polarization angle with a period of $P_{\mathrm{pol}} \approx 46 \pm 6 \min$ \citepalias{GRAVITY2018}(see also \citealt{GRAVITY2020c}). The normalized Stokes components $U/I$ and $Q/I$ trace out a loop in time, which indicates the existence of an ordered magnetic field configuration. 
It is also worth noting that the observed July 22 flare exhibits a different polarization period than the other two flares. The period of its polarization swing is $p_{\mathrm{pol}} \approx 57 \pm 8$ min, which is almost twice of the orbital period.
The highly ordered polar magnetic field in our model can indeed result in a polarization angle swing, but the oscillation frequency of stokes $Q$ and $U$ components are substantially higher, with a period of $\sim$ 20 mins for all three models, just a half of the orbital period. 
The overall evolution of $Q/I$ and $U/I$ for the two flux rope models are very similar; they trace nearly two full loops with same amplitude in the $Q-U$ diagram.
The evolution pattern of polarized emission for the flare loop is slightly different. They exhibit an ``outer'' and an ``inner'' loop in the $Q-U$ diagram respectively. The $Q-U$ signature will trace the outer loop when the ``hot spot'' is approaching to the observer and follow the inner loop when receding from the observer (see \citealt{Vos2022} for an explicit discussion). The forward flux loop is always moving toward the observer and backward rope always backing away from the observer hence lacking this "inner outer double-looped" feature.

Overall, we found that it is very hard to reproduce the oscillation period provided by \citetalias{GRAVITY2018} using our simple model. This is likely due to our simplified magnetic field setup taken from the solar corona mass ejection model  \citep{Lin2000}. The magnetic field in the black hole accretion system, on the contrary, should be more complicated due to the strong differential rotation and turbulence. The key caveat of our magnetic field configuration is the neglect of the toroidal component of the magnetic field for simplicity. While we believe that this simplification should not affect the dynamics of the ejected flux rope, it does significantly affect the polarization result. In our next work, we will use our GRMHD numerical simulation data of \citet{Miki2022} and adopt the magnetic field configuration obtained in the simulation to revisit this problem. 

Another quantity is the fraction of polarization. \citetalias{GRAVITY2018} showed a polarization degree of $20\%-40\%$. In our model, it can reach $60\%-70\%$ for the flux rope and $40\%-50\%$ for the flare loop, which is also related to our highly order magnetic field configuration adopted. This high polarization fraction is almost close to the limit set by the emission of power-law electrons in uniform magnetic field lines. For a population of power-law distributed electron with uniform magnetic field lines, the degree of polarization from synchrotron radiation is $\Pi=(p+1)/(p+7/3)$, which should be $\sim 75\%$ if $p=2.5$ or $3.5$ as in the case of our model \citep{Rybicki1986}. 
It is naturally expected that the involving of turbulence in more realistic situations can change the local magnetic field configuration to some extent, and result in the field reversal in small scale, which will decrease the polarization fraction.
In addition, external Faraday rotation and conversion can lead to depolarization when the light travels through magnetized plasma, which is not taken into consideration either. 

\section{Discussion}\label{sec:discussion}

\subsection{Model Parameter Exploration}

In our fiducial model, the backward and forward flux ropes have similar dynamical characteristics and non-thermal electrons, and thus should show similar intrinsic emissions. However, the backward blob is apparently dimmer on the camera screen than the forward one. As we have stated above, we suspect this discrepancy mainly results from  Doppler beaming and gravitational lensing.
To diagnose their respective roles, we have performed a series of ray-tracing runs by varying the viewing  angle, position and velocity of the flux rope while keeping other parameters fixed. For the inclination angle $i>90$ chosen in this work, the observer is thus located on the lower side of the equatorial plane (refer to the left plot of Figure \ref{fig:schematic}).  

\begin{figure}
	\includegraphics[width=\columnwidth]{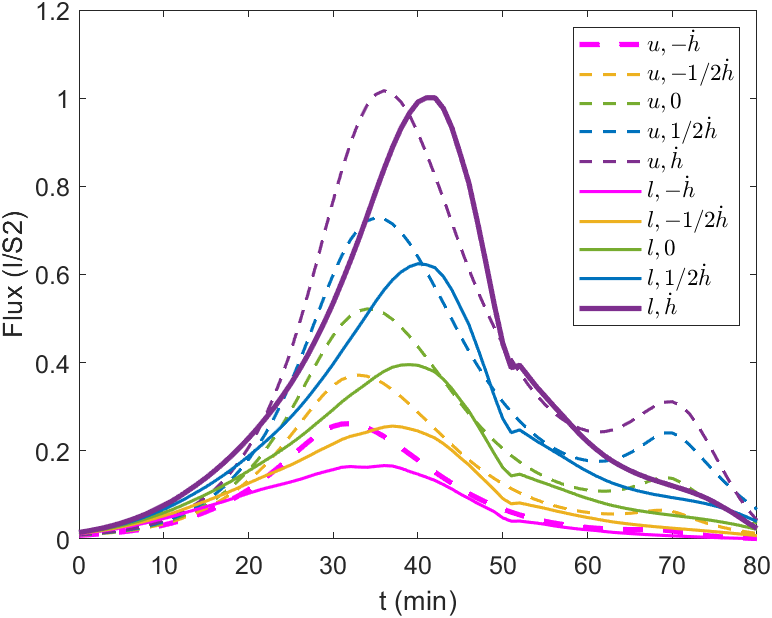}
    \caption{Light curves produced by the flux ropes with different locations and velocities. $u (l)$ denotes that the flux rope is on the upper (lower) side of the equatorial plane, $\dot{h}(-\dot{h})$ denotes that the flux rope is moving toward (away) from the observer. The initial $\phi_0=285^{\circ}$ and $\Omega=270$ for both cases. Note that for our choice of inclination angle, the camera is on the lower side of the equatorial plane. Only combinations ($u,-\dot{h}$) (backward blob) and ($l,\dot{h}$) (forward blob) are physically plausible, while others are to explore the role of Doppler boosting and gravitational lensing effect.} 
    \label{fig:7}
\end{figure}

To test the beaming effect driven by radial velocity, the radial velocity is manually assigned while the height of the blob $h$ and its angular velocity $\Omega_{\mathrm{rp}}$ is calculated as before. The sign of radial velocity is defined to be positive if the blob is moving toward us and negative if away from us.
The effect of different radial velocities is shown in Figure \ref{fig:7} with different lines (dotted lines for the backward rope and solid lines for the forward rope). By comparing with the case with no radial velocity ($\dot{h}=0$), we see that Doppler effects can boost or de-boost the intrinsic flux by a factor of 2 for our fiducial model ($\dot{h}\sim0.3-0.4c$). 

By comparing the light curves with $\dot{h}=0$ but located at two sides of the disk plane, we find that the gravitational lensing effect makes the blob behind the black hole have a greater flux. We will see this is also true for the blob with different velocities if we compare the solid and dotted lines with the same colors, even though they are now mixed with Doppler effect. The influence of lensing effect in our model is in general not so significant compared to the beaming effect. We suspect that this is because (i) the outward radial velocity is close to be relativistic and (ii) the projected position of the blob on the sky plane is relatively far away from the black hole. The former enhances the beaming effect while the latter weakens the lensing effect.

\begin{figure}
	\includegraphics[width=\columnwidth]{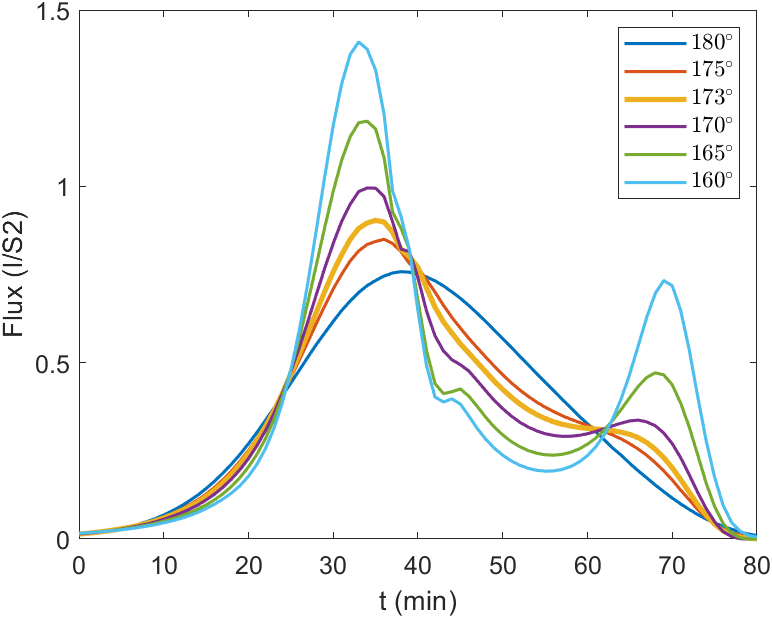}
	\includegraphics[width=\columnwidth]{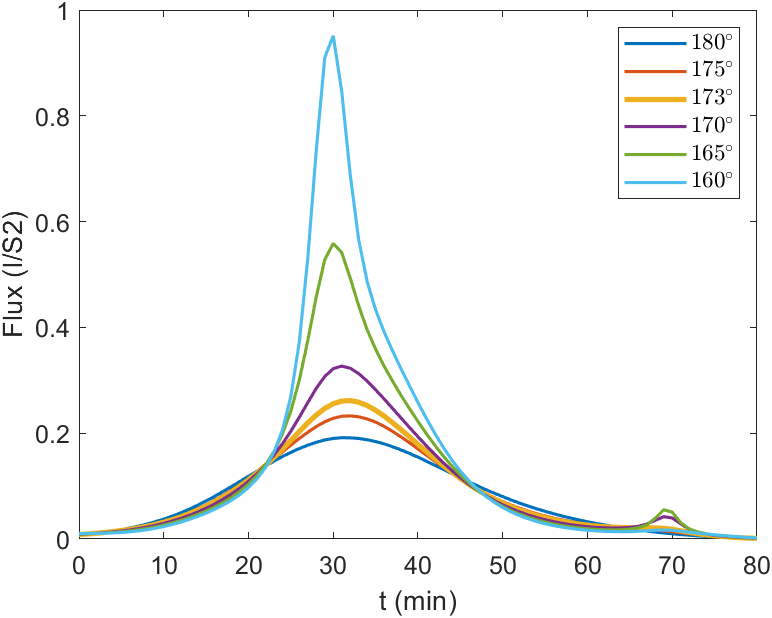}
    \caption{Light curves produced by the flux ropes with different inclination angles. The upper (lower) panel shows the forward (backward) flux rope. When the viewing angle becomes larger ($180^{\circ}-i$ is larger), the flux enhancement becomes more significant, and an apparent "double peak" structure due to beaming effect  appears.}
    \label{fig:lc_incl}
\end{figure}

To diagnose the influence of the inclination angle $i$, we perform a series of ray-tracing runs for the flux rope.
Figure \ref{fig:lc_incl} shows the light curves of the forward (upper panel) and backward (lower panel) flux ropes respectively with various values of inclination angle $i$. For both cases, increasing the value of $180^{\circ}-i$ will lead to the increase of the peak flux.  Furthermore, depending on the model parameters, a second peak could appear in the light curve. The appearance of the first peak is due to the evolution of electron acceleration during the magnetic reconnection, while the appearance of the second peak is due to the Doppler effect. These two effects are not necessarily synchronous. This remarkable double-peaked feature in the light curve is detected in the July 28th flare. Our results are comparable to the observations in the sense of the ratio of the two peak fluxes and their time interval. We note that \citet{Ball2021} also presented a model to explain the presence of the two peaks, although their physical mechanism is different from ours.  

\begin{figure}
	\includegraphics[width=\columnwidth]{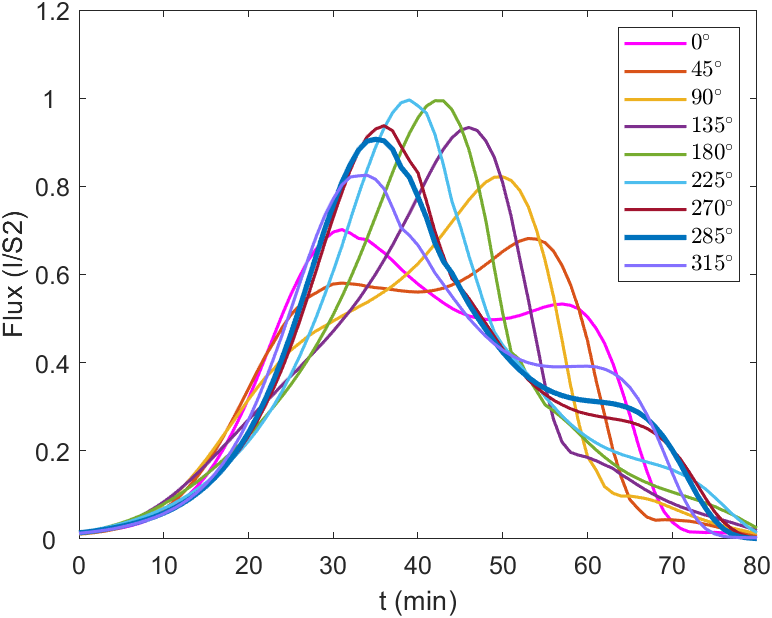}
	\includegraphics[width=\columnwidth]{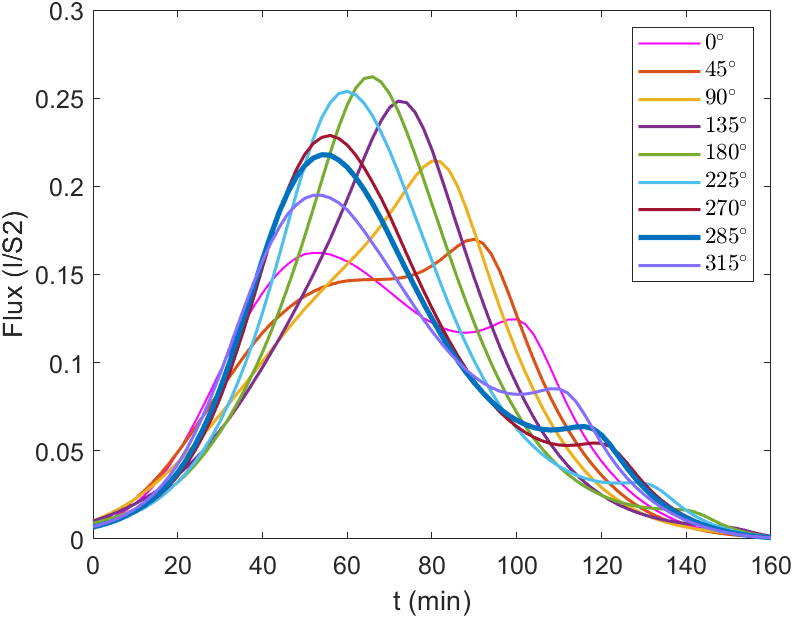}
    \caption{Light curves produced by the forward (upper panel) and backward (lower panel) flux ropes with different $\phi_0$. The two thick lines denote the fiducial model. An interesting feature is the appearance of a  second peak in the light curve for a certain range of $\phi_{0}$.}
    \label{fig:9}
\end{figure}

Another interesting dependence of light curve pattern is on the initial azimuthal angle $\phi_0$. We show in Figure \ref{fig:9} the light curves of forward (upper panel) and backward (lower panel) flux ropes with different $\phi_0$. The overall variation of the light curves for the two models is similar. The time when the most significant enhancement through beaming occurs varies with $\phi_0$. Therefore, the overall profile will be sharp if the time when the external enhancement is strongest is coincident with the time when the intrinsic radiation reaches its peak, otherwise it will be flat if the two epochs are asynchronous.  

\begin{figure}
	\includegraphics[width=\columnwidth]{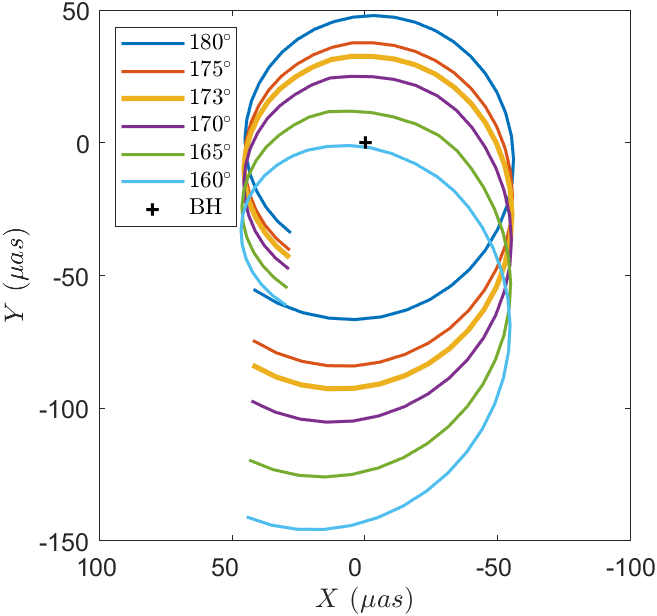}
	\includegraphics[width=\columnwidth]{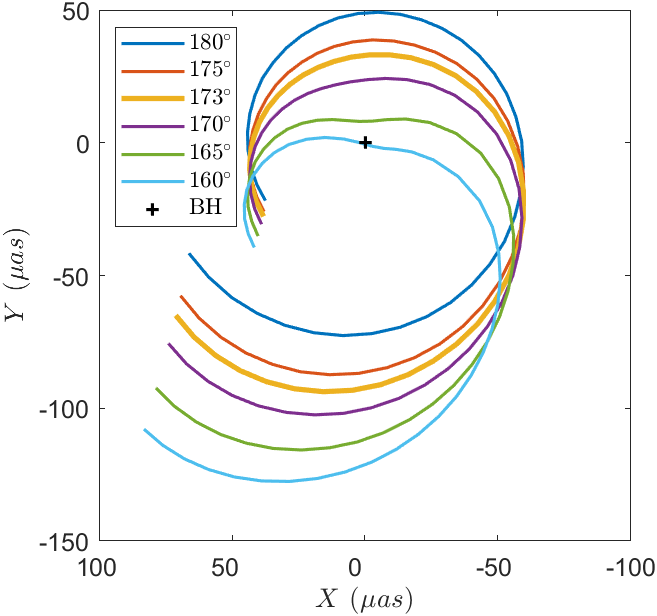}
    \caption{The centroid motion of the forward (upper panel) and backward (lower panel) moving flux ropes with different inclination angles. We only present the trajectory within 40 min during the flaring epochs.}
    \label{fig:13}
\end{figure}

Lastly, we present different centroid trajectories by changing the inclination angle $i$ for both flux rope models. The results are shown in Figure \ref{fig:13}. 
As expected, the projected trajectory varies from close to a circular pattern to a helix as the inclination angle $i$ decreases from $180^\circ$ to $90^\circ$. This is almost the case for the forward moving flux rope where the light bending effect is less important (upper panel of Figure \ref{fig:13}). The situation for the backward moving rope is more complicated. Strong gravitational lensing makes the images of the hot spot on the camera appear as two bright spots on each side of the black hole rather than a single one (see Figure \ref{fig:12}). The flux-weighted centroid thus shows a slightly smaller centroid projected distance as shown in the lower panel of Figure \ref{fig:13}. 

\subsection{Comparison with Previous Works}
\label{comparison}

By contrast to the previous works (\citetalias{Yuan2009} and \citetalias{Li2017}), the flux rope now is ejected  
along a fixed conical angle $\theta_{\mathrm{B}}$ rather than along a cylinder above the equatorial plane as we have described above (see right panel of Figure \ref{fig:schematic}). The new scenario agrees well with the flux rope trajectory found in the 3D GRMHD simulation of \citet{Miki2022}\footnote{\citet{Miki2022} find that collimation effect exists for the ejected flux rope, i.e, the corresponding value of $\theta$ in spherical coordinate of the flux rope gradually decreases with its outward propagation. Our definition of $\theta$ is different from the standard definition in a spherical coordinate because the flare loop is not located at the position of the black hole (refer to Figure \ref{fig:schematic}). Because of this reason, our constant-$\theta_B$ model effectively has taken into account the collimation effect. }. A similar ejected flux rope scenario has been studied by \citet{Younsi2015} and \citet{Ball2021}. By analogy with the same CME model, \citet{Younsi2015} considered the case where the flux rope is ejected out but confined on a cylinder surface. Therefore, we expect that their model is difficult to be compatible with the super-Keplerian motion of the observed hot spot.
Our flux rope trajectory is also  different from work of \citet{Ball2021}. They considered a phenomenological hot spot model and the spot is moving along a fixed $\theta_0$ in spherical coordinates with the generatrix of the cone points towards to the central black hole. 

There are also some GRMHD simulations aimed at interpreting the flares from \sgra. In \citet{Porth2021}, the magnetically dominant plasmoids are embedded within the accretion flow, and the motion of these plasmoids is thus substantially sub-Keplerian, which is in tension with the GRAVITY observations. \citet{Petersen2020} also focus on the accretion flow rather than ejection. Their model can reproduce the flux and spectrum of the IR flares, but fails to explain the centroid motion of the hot spot observed by GRAVITY. In some GRMHD simulation works, it has been found  that the plasmoids can be be ejected out from the accretion flow \citep{Nathanail2022,Ripperda2022}, but a quantitative calculation of the radiative transfer and comparison with GRAVITY observations in terms of the centroid motion and flare light curves usually lacked in these works. 

\subsection{Implications to the observed quasi-periodic variability of the NIR flares from Sgr A*}

While the present paper focuses on the interpretation of the GRAVITY observational results, it is interesting to discuss other infrared  observations to the flares of \sgra. Among these observations, one prominent result we think presents an important constraint on the nature of Sgr A* flare is the periodicity of the flares detected  by the high-resolution infrared observations of \citet{Genzel2003}. They have performed the power spectrum analysis to two light curves lasting $\sim 130$ minutes and found a significant peak with a time period of  $\sim 17$ minutes. The periodicity is confirmed later by other infrared observations \citep{2006A&A...455....1E,2007MNRAS.375..764T,2010RvMP...82.3121G}\footnote{ We note that some observations do not find such a periodic component from the light curves and the variability is fully consistent with the red-noise \citep[e.g.,][]{2009ApJ...691.1021D}. As pointed out by \citet{2010RvMP...82.3121G}, the apparent discrepancy seems to be the
question whether it is reasonable for the observer to select certain parts of the light curve to analyze the significance of the periodicity, which may be washed out in the noise-like variable state of longer
time series.}, although the period changes somewhat from case to cases. This period is consistent with the orbital period at 3–5$r_g$, assuming the accretion flow in Sgr A* is described by a magnetically arrested disk (MAD).

Such a quasi-periodic variability is hard to be explained by the orbital motion of a single hot spot. Our detailed modeling shown in the present paper indicates that the flare  due to a single hot spot usually lasts for less than one orbital timescale. The physical reasons are: (i) As we have discussed above, synchrotron cooling timescale for these electrons is very short in the high magnetic field environment. So the duration of the flare is determined by the magnetic reconnection, which cannot last for too long either. (ii) The ejected blob can rush out to a very large distance in a short time, which causes the magnetic field embedded in the blob drops rapidly. Therefore, the magnitude of the radiation would decline and even becomes weaker than the threshold of the detection limit. 

Instead, we suggest that the periodic formation of the flux ropes found in the numerical simulations of \citet{Miki2022} may be responsible for the quasi-periodic variability. It is found that flux ropes are formed periodically, with the period being the local orbital timescale. Specifically, they found that the flux ropes formed within $10-15\ r_g$ are hard to be ejected out but stay within the accretion flow and finally advected into the black hole horizon. So it is predicted that, the flares with short period of variability as found in \citet{Genzel2003} should not be accompanied by blob ejections, thus their trajectories projected on the sky should not become larger with time. 

\section{Summary}
\label{sec:6}

High-resolution GRAVITY observations have resolved the infrared flares of Sgr A* and a compact “hot
spot” was detected orbiting around the black hole with a high speed.  Among other features, two prominent results obtained in the observations are that: 1) the trajectory of the hot spot gradually increases with time; and 2) the radius of the circular trajectory combined with the orbital time indicates that the hot spot is rotating with super-Keplerian speed \citep{GRAVITY2018}. These results present important constraint on the nature of flares in \sgra. 

\citet{Yuan2009} has proposed a ``CME'' model to explain the flares by analogy with the coronal mass ejection in the Sun. The basic scenario is that, magnetic reconnection at the surface of the accretion flow results in the formation of flux ropes, which are then ejected out with high velocity. The scenario has been confirmed by our recent GRMHD simulations \citep{Miki2022}. Electrons accelerated in the reconnection current sheet escape to the  regions of the flare loop and of the ejected flux rope, whose radiation are responsible for the flares. The first application of the model to the interpretation of Sgr A* was presented in \citet{Li2017}. In that work, the dynamics of the ejected blobs, the energy distribution of accelerated nonthermal electrons and their radiation were calculated and compared with the light curve and spectrum of infrared and X-ray flares. It was found that both the radiation from the ejected blob and flare loop provided the comparable contribution to the observed flares.  

In the present paper, we further develop the \citet{Li2017} model by also taking into account the toroidal motion of the ejected blobs in addition to its poloidal motion, which was neglected in \citet{Li2017} for simplicity (refer to Figure \ref{fig:schematic} for the schematic figure). 
In our model, the ejected flux rope follows a spiral motion roughly confined in the surface of a cone. Therefore the distance of the hot spot from the black hole  in the projected plane gradually becomes larger with time. Moreover, as suggested by the numerical simulations of \citet{Miki2022}, the angular velocity of the blob roughly remains the same value as its launching location as long as it is not too far away from this launching location. Thus, with the increase of radius in the spiral motion, its rotation becomes  super-Keplerian, fully consistent with  \citet{GRAVITY2018}. We consider the electron acceleration by reconnection and the evolution of the energy distribution of accelerated electrons. The radiation calculation is elaborated by  carrying out fully general relativistic ray-tracing radiative transfer calculations, which is another main improvement compared to \citet{Li2017}. 

We have considered three components as the candidate model of Sgr A* flares, namely the forward and backward motion flux ropes and the flare loop. We find that the radiation from the ejected blobs will dominate over that from the flare loop. The forward flux rope can well reproduce both the light curve and centroid motion for the July 22 flare reported by \citet{GRAVITY2018}. Based on the trajectory of the hot spot, we can rule out the flare loop model due to a much smaller, and nearly circular pattern of the centroid motion.  But we cannot fully rule out the backward flux rope since this needs a thorough parameter survey, which is out the scope of the current work.  In addition, the flux rope models can generate double-peak structure in the light curve by adjusting the inclination angle or the initial azimuthal angle where the rope is launched.  This could be responsible for another flare observed by GRAVITY on July 28 \citep{GRAVITY2018}. 

One main caveat of our current model is that the magnetic field configuration is adopted from solar coronal mass ejection model, which is in order to obtain an analytical solution of the dynamics of the ejected flux rope. Such a highly ordered magnetic field configuration however overestimates the degree of polarization and produces a shorter polarization swing period. We expect that the future modeling by making use of the GRMHD simulation data with a more reasonable field configuration would alleviate this discrepancy.

\section*{Acknowledgements}
This work is supported in part by the Natural Science Foundation of China (grants 12133008, 12192220, and 12192223). We thank the beneficial discussions with Hai Yang, Haocheng Zhang, Antonios Nathanail, and useful comments from the referee. The calculations have made use of the High Performance Computing
Resource in the Core Facility for Advanced Research Computing
at Shanghai Astronomical Observatory. 

\section*{Data availability}
The data underlying this article will be shared on reasonable request to the corresponding author.

\bibliographystyle{mnras}
\bibliography{ms_v1.bib}{}

\appendix
\section{Transformation of velocity and Magnetic field}
\label{Appendix}


Since our magnetohydrodynamic model is calculated under the classic Newtonian frame, we have to transform the quantities we deduced into the Kerr metric in order to calculate the radiative transfer by using GRTRANS. The spacetime geometry around a rotating black hole can be described by Boyer-Lindquist (BL) coordinate, the form of which is
\begin{eqnarray}
\mathrm{d} s^2 & = & g_{\alpha \beta} \mathrm{d} x^\alpha \mathrm{d} x^\beta \nonumber \\
& = & -\alpha^2 \mathrm{~d} t^2+\gamma_{i j}\left(\mathrm{~d} x^i+\beta^i \mathrm{~d} t\right)\left(\mathrm{d} x^j+\beta^j \mathrm{~d} t\right),
\end{eqnarray}
where $i, j=r,\ \theta,\ \phi$, and  $\alpha=\sqrt{\frac{\sum \Delta}{A}}$ is the lapse function, $\beta^\phi=-\omega$ is the shift vector, $\gamma_{r r}=\frac{\Sigma}{\Delta}$, $\gamma_{\theta \theta}=\Sigma$, $\gamma_{\phi \phi}=\frac{A \sin ^2 \theta}{\Sigma}$ are the spatial component of the metric while other elements are zero. Here we choose the geometric unit where $G=c=1$ and therefore $\Sigma=r^2+a^2 \cos ^2 \theta, \Delta=r^2-2 M r+a^2$, $A=\left(r^2+a^2\right)^2-a^2 \Delta \sin ^2 \theta$, and $a$ is the angular momentum of the black hole per unit mass. The angular velocity due to the frame dragging is $\omega=2 M a r /A$.

To transform the classic 3-velocity to relativistic 4-velocity. We first decompose the poloidal velocity $\dot{h}$ of the flux rope into spherical coordinate and obtain $v^{\hat{r}}$ and $v^{\hat{\theta}}$.

We then approximate these velocity components are measured by the zero angular momentum observer (ZAMO) in the locally non-rotatating frame (LNRF, \citealt{Bardeen1972})
 who has four-velocity
\begin{equation}
u_{\mathrm{LNRF}}=\left(u_{\mathrm{LNRF}}^t, 0, 0,\omega u_{\mathrm{LNRF}}^t\right).
\end{equation}

Therefore, the 3-velocity of the flux rope with respect to the LNRF can be approximately described by 
($v^{\hat{r}}$, $v^{\hat{\theta}}$,$v^{\hat{\phi}}$), where
\begin{equation}
v^{\hat{\phi}}=\frac{A}{\Sigma \Delta^ {1 / 2}} \widetilde{\Omega} \quad ; \quad \widetilde{\Omega} \equiv \Omega_{\mathrm{rp}}-\omega,
\end{equation}
and the flux rope 4-velocity with respect to the LNRF is 
\begin{equation}
u^{\hat{\alpha}}=(u^{\hat{t}},u^{\hat{r}},  u^{\hat{\theta}},u^{\hat{\phi}}),
\end{equation}
where $u^{\hat{t}}=\gamma=\left(1-v^2\right)^{-1 / 2}$ with $v^2=v^{\hat{i}}v_{\hat{i}}$ and $u^{\hat{i}}=\gamma v^{\hat{i}}$ as in special relativity.

Finally we transform these velocity components from LNRF to BL coordinate frame 
by using orthonormal tetrads carried by LNRF \citep{Bardeen1972}:

\begin{equation}
\begin{aligned}
&e^{\mu}_{\ \hat{t}}=\sqrt{\frac{A}{\Sigma \Delta}} (1,0,0,\frac{2Mar}{A}), \\ &e^{\mu}_{\ \hat{r}}=\sqrt{\frac{\Delta}{\Sigma }} (0,1,0,0), \\
&e^{\mu}_{\ \hat{\theta}}=\sqrt{\frac{1}{\Sigma} }(0,0,1,0), \\
&e^{\mu}_{\ \hat{\phi}}=\sqrt{\frac{\Sigma}{A}} \frac{1}{\mathrm{sin} \theta}(0,0,0,1).
\end{aligned}
\end{equation}

The 4-velocity of the flux rope in the BL frame can then be written as
\begin{equation}
u^{\mu}=e^{\mu}_{\hat{\alpha}}u^{\hat{\alpha}}.
\end{equation}

The components of the magnetic field in the classic frame $B$ is described by Eq. \ref{eq:2}. As same as velocity, we approximate that they are the quantities observed by the ZAMO and the magnetic field in the LNRF can thus be written as $B_{\mathrm{LNRF}}= (0,B^{\hat{r}},B^{\hat{\theta}},0)$, where we approximate $B^{\hat{t}}=B^{\hat{\phi}}=0$. The square of the field magnitude is given by $B^2=B^{\hat{i}}B_{\hat{i}}$. The BL coordinate frame contravariant components of the magnetic field therefore can be written as \citep{GRAVITY2020c}

\begin{equation}
\begin{aligned}
B^t & =-C B^\theta ; \\
B^r & =\delta_c B_\theta=B_\theta \delta_{\rm LNRF} / r ; \\
B^\theta & =B\left(g_{t t} C^2+g_{r r} \delta_c^2+g_{\theta \theta}\right)^{-1 / 2} \\
B^\phi & =0
\end{aligned}
\end{equation}
with 
\begin{equation}
C \equiv \frac{\delta_c g_{r r} u^r+g_{\theta \theta} u^\theta}{g_{t t} u^t+g_{t \phi} u^\phi} \quad ; \quad\delta_c=\delta_{\rm LNRF} / r, 
\end{equation}
where $\delta_{\rm LNRF}=\frac{B^{\hat{r}}}{B^{\hat{\theta}}} $ is the ratio of radial and polar magnetic field components in the LNRF.

The field can then be transformed to the orthonormal
frame comoving with the fluid by using tetrads \citep{Dexter2016}
\begin{equation}
\begin{aligned}
& \boldsymbol{e}_{(t)}^\mu=u^\mu, \\
& \boldsymbol{e}_{(r)}^\mu=\left(u_r u^t,-\left(u_t u^t+u_\phi u^\phi\right), 0, u_r u^\phi\right) / N_r, \\
& \boldsymbol{e}_{(\theta)}^\mu=\left(u_\theta u^t, u_\theta u^r, 1+u_\theta u^\theta, u_\theta u^\phi\right) / N_\theta, \\
& \boldsymbol{e}_{(\phi)}^\mu=\left(u_\phi, 0,0,-u_t\right) / N_\phi,
\end{aligned}
\end{equation}
where 
\begin{equation}
\begin{aligned}
& N_r^2=-g_{r r}\left(u_t u^t+u_\phi u^\phi\right)\left(1+u_\theta u^\theta\right), \\
& N_\theta^2=g_{\theta \theta}\left(1+u_\theta u^\theta\right), \\
& N_\phi^2=-\left(u_t u^t+u_\phi u^\phi\right) \Delta \sin ^2 \theta.
\end{aligned}
\end{equation}

Components of the magnetic field in the BL coordinate frame are transformed to the fluid-frame by
\begin{equation}
A^{(\beta)}=\eta^{(\alpha)(\beta)} g_{\mu \nu} e_{(\alpha)}^\mu A^\nu
\end{equation}
where $\eta^{(\alpha)(\beta)}$ is the Minkowski metric.

\bsp	
\label{lastpage}

\end{document}